\documentclass[11pt,aps, prd, superscriptaddress,nofootinbib]{revtex4-2}

\usepackage[utf8]{inputenc}
\usepackage{enumitem,amssymb}
\usepackage{graphicx}
\usepackage[export]{adjustbox}
\usepackage[dvipsnames]{xcolor}    
\usepackage{fontawesome}

\usepackage{amsmath,amsfonts,bm,slashed,array,multirow,xspace}
\usepackage{setspace}

\definecolor{xlinkcolor}{cmyk}{1,1,0,0}
\usepackage{url}
\usepackage[
 colorlinks=true,    
 linkcolor=xlinkcolor,     
 citecolor=xlinkcolor,     
 filecolor=xlinkcolor,  
 urlcolor=xlinkcolor,      
 final=true
]{hyperref}

\allowdisplaybreaks

\makeatletter
\g@addto@macro\bfseries{\boldmath}
\makeatother

\newcommand\snowmass{\begin{center}\rule[-0.2in]{\hsize}{0.01in}\\\rule{\hsize}{0.01in}\\
\vskip 0.1in Submitted to the  Proceedings of the US Community Study\\ 
on the Future of Particle Physics (Snowmass 2021)\\ 
\rule{\hsize}{0.01in}\\\rule[+0.2in]{\hsize}{0.01in} \end{center}}

\def\vect#1{\boldsymbol{#1}}

\begin{document}

\snowmass
\vspace{-15pt}
\title{Snowmass White Paper: Light Dark Matter Direct Detection\\at the Interface With Condensed Matter Physics}

\author{Andrea~Mitridate}
\author{Tanner~Trickle}
\affiliation{Walter Burke Institute for Theoretical Physics,\\ California Institute of Technology, Pasadena, CA 91125, USA}

\author{Zhengkang~Zhang}
\affiliation{Department of Physics, University of California, Santa Barbara, CA 91106, USA}

\author{Kathryn~M.~Zurek}
\affiliation{Walter Burke Institute for Theoretical Physics,\\ California Institute of Technology, Pasadena, CA 91125, USA}

\begin{abstract}
    Direct detection experiments for light (sub-GeV) dark matter are making enormous leaps in reaching previously unexplored theory space. 
    The need for accurate characterizations of target responses has led to a growing interplay between particle and condensed matter physics.
    This white paper summarizes recent progress on direct detection calculations that utilize state-of-the-art numerical tools in condensed matter physics and effective field theory techniques. 
    These new results provide the theoretical framework for interpreting ongoing and planned experiments using electronic and collective excitations, and for optimizing future searches.
\end{abstract}
\maketitle

\section{Executive Summary}

Direct detection of dark matter (DM) has undergone a dramatic expansion of scope in recent years.
Theoretical development of new DM models together with experimental advances on cryogenic superconducting calorimeters have given impetus to new proposals to search for light (sub-GeV) DM.
For DM in the sub-GeV regime, conventional searches based on nuclear recoil lose sensitivity due to kinematic mismatch, whereas electronic excitations and collective excitations in condensed matter systems provide alternative pathways to discovery. 

The theory community has been actively pursuing novel detection ideas along these lines. 
A variety of target systems and detection strategies have been investigated, including electronic excitations in noble gas atoms~\cite{Essig:2011nj, Catena2020, Graham:2012su, Lee:2015qva, Essig:2017kqs, DarkSide:2018ppu, XENON:2019gfn, XENON:2020rca} and dielectric crystals~\cite{Griffin2021, Mitridate2021, Griffin2020, Trickle2020, Griffin2021a, Marshall2021, Kurinsky2019, Essig2016, Derenzo2017, Knapen2021, Hochberg2021}, superconductors~\cite{Hochberg:2015pha, Hochberg:2016ajh}, graphene~\cite{Hochberg:2016ntt}, Dirac materials~\cite{Hochberg2018, Coskuner2021, Geilhufe2020}, aromatic organics~\cite{Blanco2020, Blanco2021}, via the Migdal effect~\cite{Kahn:2020fef,Knapen:2020aky}, molecular dissociation or excitation~\cite{Essig:2016crl,Arvanitaki:2017nhi,Essig:2019kfe}, defect production~\cite{Budnik:2017sbu,Rajendran:2017ynw}, collective excitations in superfluid helium~\cite{Schutz:2016tid,Knapen:2016cue}, phonon~\cite{Knapen:2017ekk,Griffin:2018bjn} and magnon~\cite{Trickle:2019ovy} excitations in crystals, and axion quasiparticles in antiferromagnetic topological insulators~\cite{Marsh:2018dlj,Schutte-Engel:2021bqm,Chigusa:2021mci}.
Meanwhile, several ongoing experiments, including XENON~\cite{XENON:2019gfn, XENON:2020rca}, DarkSide~\cite{DarkSide:2018ppu}, DAMIC~\cite{DAMIC:2015znm, DAMIC:2019dcn, Settimo:2020cbq}, SENSEI~\cite{Tiffenberg:2017aac, Crisler:2018gci, SENSEI:2019ibb, SENSEI:2020dpa}, SuperCDMS~\cite{SuperCDMS:2014cds, SuperCDMS:2015eex, SuperCDMS:2016wui, SuperCDMS:2017nns, SuperCDMS:2018gro, SuperCDMS:2018mne, SuperCDMS:2020ymb}, CDEX~\cite{CDEX:2017kys} and EDELWEISS~\cite{EDELWEISS:2018tde, EDELWEISS:2019vjv, EDELWEISS:2020fxc}, are already conducting searches with electronic excitations readout. 
R\&D for phonon readout is also underway, pioneered by SPICE and HeRALD experiments as part of the TESSERACT project~\cite{tesseract}. 
These new developments present exciting opportunities to cover well-motivated, but yet unexplored, DM theory space.
See Ref.~\cite{Kahn:2021ttr} for a recent review.

Crucial to the success of this ongoing program of light dark matter direct detection is to have 
accurate characterizations of target response in all relevant kinematic regimes. 
This involves interdisciplinary research to bridge particle physics models and condensed matter calculations. 
On the condensed matter side, density functional theory (DFT) plays a key role in \textit{ab initio} calculations of material properties, and data-driven approaches have also been explored that provide complementary information in some cases~\cite{Knapen2021, Hochberg2021, Knapen2022}. 
On the particle physics side, effective field theory (EFT) provides the necessary framework and tools for both systematically exploring the DM theory space and matching onto target responses~\cite{Catena2020,Trickle:2020oki,Catena2021}. 
Accurate characterization of background for low-threshold detectors is also essential, and there has been rapid progress on this subject recently~\cite{Du:2020ldo,Berghaus:2021wrp}.

In what follows, we highlight some of the key achievements in direct detection calculations.
\begin{itemize}
	\item For electronic excitations, signal rate predictions have been significantly improved with a combination of DFT and analytic tools to more accurately model the electronic states in dielectric crystals~\cite{Mitridate2021, Chen2022, Griffin2021}.
	\item For collective excitations, a repository of phonon calculations containing more than twenty target materials has been established~\cite{Griffin:2019mvc,Coskuner:2021qxo,demo}, which provides theory input on optimization of experimental target choice. Also, an EFT framework for phonon and magnon rate calculations has been developed~\cite{Trickle:2020oki}, allowing for interpretations of near future experimental progress for broad classes of DM models.
	\item Numerical tools, including \textsf{EXCEED-DM}~\cite{Trickle2022a, Griffin2021} and \textsf{PhonoDark}~\cite{phonodark}, have been developed to facilitate these calculations.
\end{itemize}
Light DM direct detection will play an increasingly important role in the DM search program in the next decade. 
We look forward to continued interdisciplinary collaboration to move the field forward.

\section{Electronic Excitations}

\begin{figure}
    \centering
    \includegraphics[width=0.25\textwidth, valign=c]{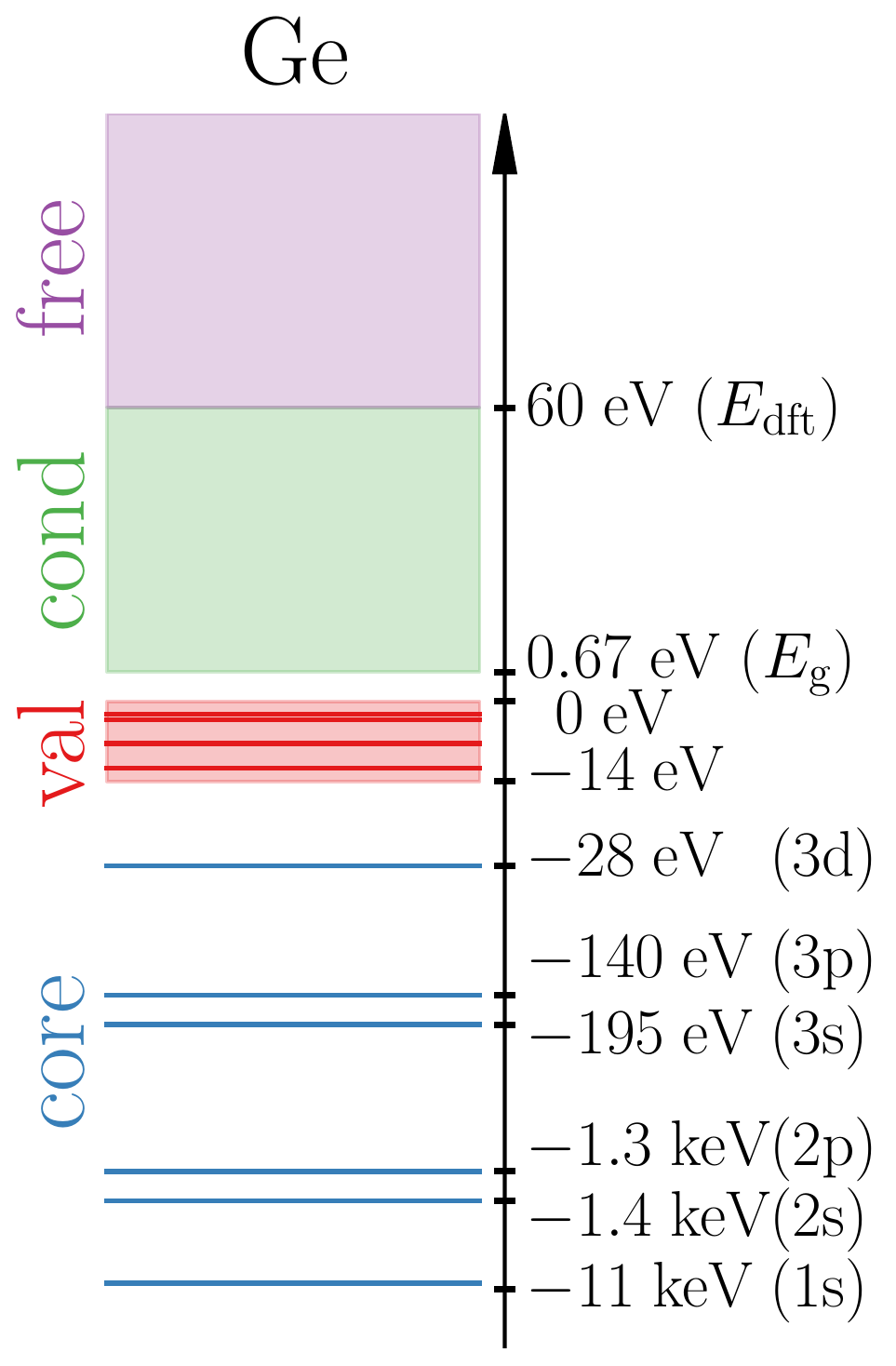}
    \includegraphics[width=0.7\textwidth, valign=c]{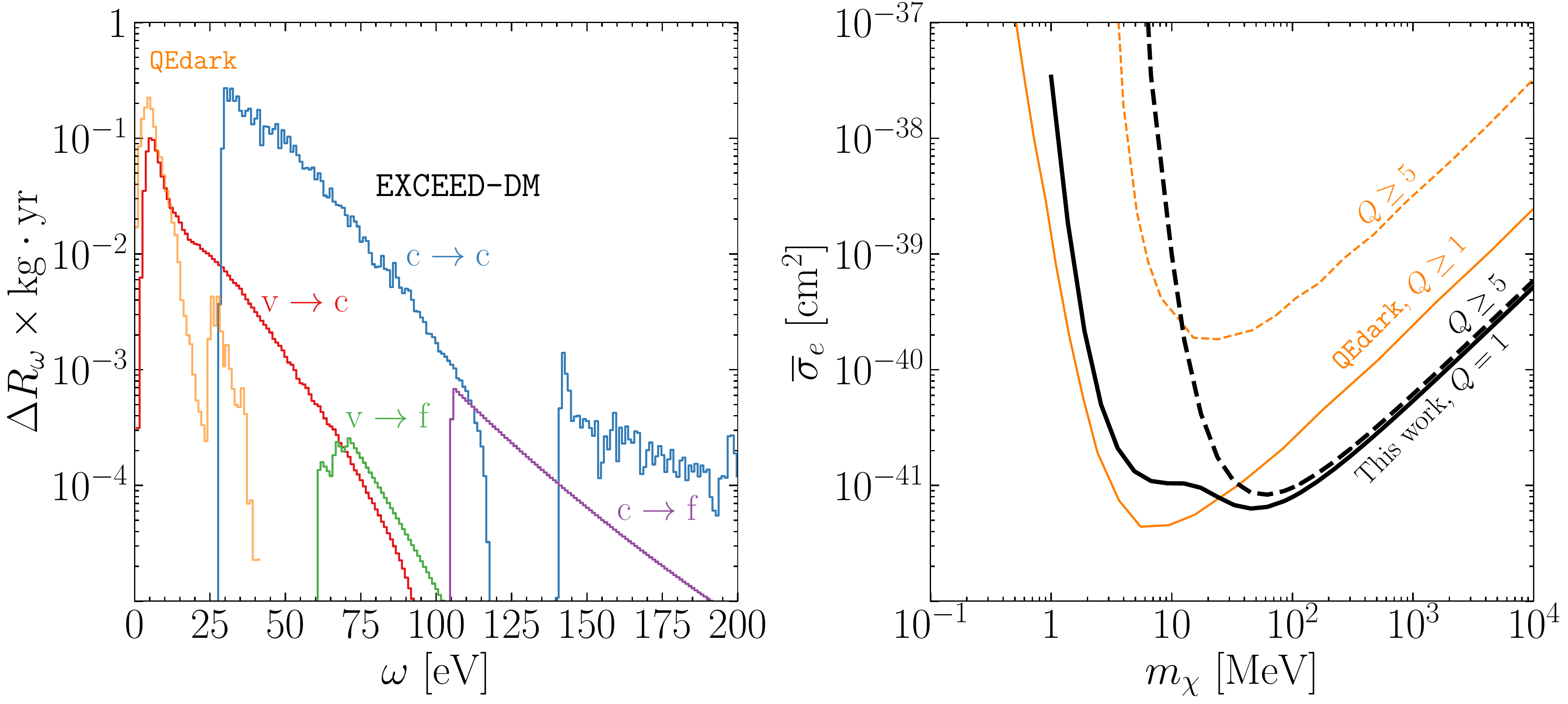}
    \caption{(Reproduced from Ref.~\cite{Griffin2021}.) \textbf{Left}: Illustration of electronic states in a Ge target. \textbf{Center}: Binned number of events in a Ge target with a kg-yr exposure, assuming a DM mass of $\mathrm{GeV}$ and a heavy mediator. Contributions from the different transition types, valence to conduction ($\mathrm{v} \rightarrow \mathrm{c}$), core to conduction ($\mathrm{c} \rightarrow \mathrm{c}$), valence to free ($\mathrm{v} \rightarrow \mathrm{f}$), and core to conduction ($\mathrm{c} \rightarrow \mathrm{f}$) are shown. Lines labelled \texttt{QEdark} are previous projections from Refs.~\cite{Essig2016, Derenzo2017} considering unscreened, valence to conduction transitions. \textbf{Right}: 95\% C.L. (3 events) constraints on the scattering cross section assuming a kg-yr exposure in a Ge target assuming a heavy mediator. $Q$ is the number of electron-hole pairs produced in an event, and corresponds to the experimental energy threshold. For $Q = 5$ (threshold energies of $\sim 15\,\text{eV}$) we find significantly improved constraints due to the improved modelling of the high momentum components of the electronic states.}
    \label{fig:electronic_scatter_fig}
\end{figure}

Light DM can drive electronic excitations in a variety of target systems, from transitions across the band gap in semiconducting, or dielectric, crystals~\cite{Griffin2021, Mitridate2021, Griffin2020, Trickle2020, Griffin2021a, Marshall2021, Kurinsky2019, Essig2016, Derenzo2017, Knapen2021, Hochberg2021}, Dirac materials~\cite{Hochberg2018, Coskuner2021, Geilhufe2020}, or materials with spin-orbit coupling~\cite{Inzani2021, Chen2022}, to ionization from atomic targets~\cite{Essig:2011nj, Catena2020, Graham:2012su, Lee:2015qva, Essig:2017kqs, DarkSide:2018ppu, XENON:2019gfn, XENON:2020rca}, and transition between molecular orbitals~\cite{Blanco2020, Blanco2021}. All of these transitions involve an electron being kicked from an initial state to a final state with enough energy to cross a target-specific threshold. The thresholds can range from $\mathcal{O}(\text{meV})$ band gaps in targets with spin-orbit coupling, to $\mathcal{O}(\text{eV})$ in typical semiconductors, to $\mathcal{O}(10\,\text{eV})$ to ionize electrons in atomic targets. Kinematically this means these targets are sensitive to the scattering of DM with mass greater than $\mathcal{O}(\text{keV})$, and absorption of DM with mass greater that $\mathcal{O}(\text{meV})$; a well motivated but unexplored region of DM parameter space.

Currently, the direct detection experiments most sensitive to light DM are those looking for electronic excitations in semiconducting crystals such as Si and Ge (e.g., SENSEI, SuperCDMS, Edelweiss, DAMIC, CDEX). The DM interaction rate not only crucially depends on the band structure in these targets, but also the electronic wave functions, which act as a form factor relative to the free-electron calculations. Accurate modelling of these electronic states is crucial for predictions of scattering and absorption rates for general DM models. In Ref.~\cite{Griffin2021} we dramatically extended the scope and accuracy of DM-electron scattering rate calculations by including states below the valence bands, ``core" states shown in the left panel of Fig.~\ref{fig:electronic_scatter_fig}, and states above the conduction bands, ``free" states, as well as including all-electron reconstruction effects which results in more accurately modelling the high momentum components of the valence electronic states. Moreover, we included screening effects which can be important for the low momentum, low energy parts of phase space. Our results for the improved scattering calculation are shown in Fig.~\ref{fig:electronic_scatter_fig} relative to the previous best calculation, \texttt{QEdark}, which included unscreened, valence to conduction transitions without all-electron reconstruction effects. 

\begin{figure}
    \centering
    \includegraphics[width=\textwidth]{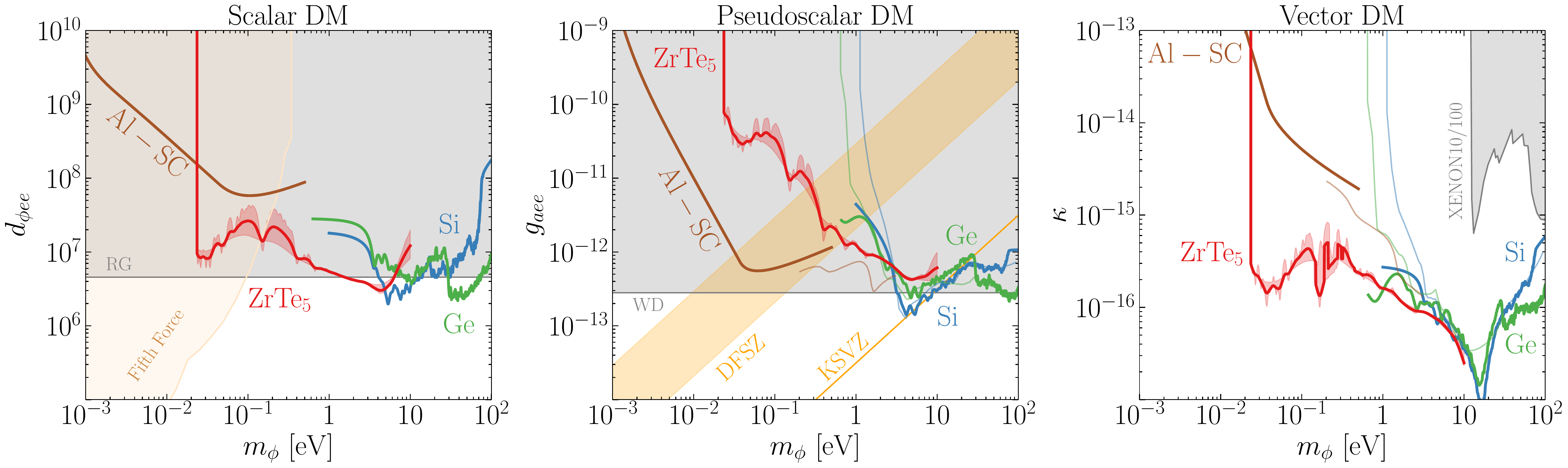}
    \caption{(Reproduced from Refs.~\cite{Chen2022, Mitridate2021}.) 95 \% C.L.\ projected constraints on scalar (left), pseudoscalar (center), and vector (right) DM models assuming no background and a kg-year exposure.}
    \label{fig:electronic_abs_fig}
\end{figure}

In addition to advancing the DM-electron scattering rate calculations, in Refs.~\cite{Mitridate2021, Chen2022} we expanded the calculation of DM-electron absorption. In targets with spin degenerate electronic eigenstates, it has been known that the absorption of vector and pseudoscalar DM can be related to optical data, which was previously thought to be the case for scalar DM as well. In Ref.~\cite{Mitridate2021}, we constructed an NR EFT for DM-electron absorption processes and showed that the scalar DM absorption rate is dominated by an operator not related to optical data. In addition, in Ref.~\cite{Chen2022} we extended this analysis to targets with spin-orbit coupling where only the vector DM can be related to optical data. Our results are shown in Fig.~\ref{fig:electronic_abs_fig}. These analyses show the importance of detailed first-principles calculations of DM-electron interactions and further extend the theory space that direct detection experiments can cover. The program computing all of the DM-electron scattering and absorption rates in general semiconducting and spin-orbit coupled materials, \texttt{EXCEED-DM}~\cite{Griffin2021, Trickle2022a}, is publicly available on Github~\href{https://github.com/tanner-trickle/EXCEED-DM}{\faicon{github}}. 

A number of advances in understanding DM-electron interaction rates have been made in parallel directions: an EFT description of electronic scattering off of atomic targets has been developed~\cite{Catena2020} which greatly expands the model space which can be probed in noble liquid experiments such as XENON1T and DarkSide. More recent work has developed a similar EFT for crystal targets~\cite{Catena2021}, however it crucially relies on the results of \texttt{QEdark} which are not sufficient for DM models with stronger $q$ dependence than the standard light mediator models (which constitutes a large class of the models within the EFT framework). Recently another approach to calculating DM-electron interaction rates has appeared, which reuses the dielectric~\cite{Knapen2021, Hochberg2021, Knapen2022} for both scattering and absorption calculations. As discussed before this approach has been used for some time for DM absorption rates, but a more general formula, including the $\mathbf{q}$ dependence of the dielectric, has been shown to be useful for DM-electron scattering. While just a reformulation of what has been previously computed, it makes the presence of screening clear. Moreover, if the dielectric were to be measured in the entire kinematic regime available to DM one could use this to set constraints. However even for the simplest targets, e.g. Si and Ge, this is a challenge, and one usually resorts to only using the static limit of the dielectric, $\epsilon(\mathbf{q} \rightarrow 0, \omega)$, and therefore inherently limited to DM models with a light mediator. On top of this the dielectric can only be used for spin-independent DM models, where the DM-electron coupling can be related to the electron-photon coupling. Therefore while useful for the simplest DM model with a light mediator, this approach is not general enough to encompass the wide range of possible DM models. The substantial amount of work being done is indicative of how important these calculations are, and further developments will only help to expand the theory space probe-able by current and upcoming experiments.

\section{Collective Excitations}

For energy depositions below the electronic band gap, collective excitations are the degrees of freedom that may couple to the DM. 
These were first considered for superfluid helium~\cite{Schutz:2016tid,Knapen:2016cue} (which was further studied in subsequent works using EFT techniques~\cite{Acanfora:2019con,Caputo:2019cyg,Caputo:2019xum,Baym:2020uos,Matchev:2021fuw}).
Later on, acoustic and optical phonons in (polar) crystals were advanced and shown to have the best experimental prospects and sensitivity to light DM~\cite{Knapen:2017ekk,Griffin:2018bjn,Kurinsky:2019pgb,Trickle:2019nya,Griffin:2019mvc,Mitridate:2020kly,Griffin:2020lgd,Coskuner:2021qxo}.
An attractive feature of crystal targets is that anisotropies in the crystal structure result in a daily modulation signal, which can be crucial for making a discovery in the presence of irreducible non-modulating background~\cite{Griffin:2018bjn,Coskuner:2021qxo}.

\begin{figure}
    \centering
    \includegraphics[width=0.7\textwidth]{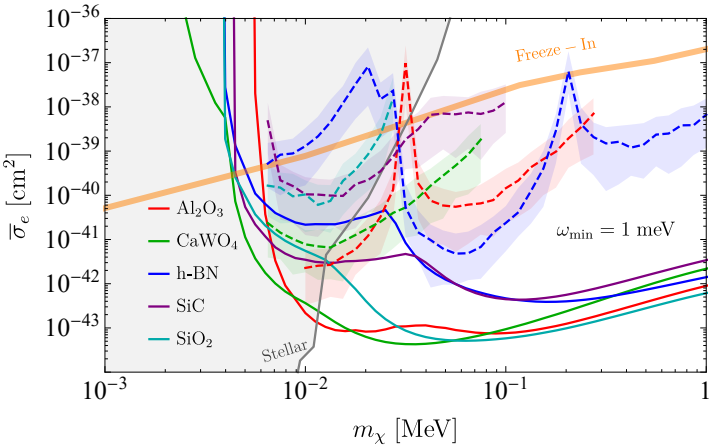}
    \caption{(Reproduced from Ref.~\cite{Coskuner:2021qxo}.) Projected constraints on the dark matter interaction cross section in the dark photon mediator model from a few anisotropic crystal targets. Solid curves are the 95 \% C.L.\ constraints (3 events), and dashed curves indicate when a daily modulation will be statistically significant. All projected constraints assume one kg-yr exposure and no background.}
    \label{fig:phonon}
\end{figure}

In an effort to provide theory input for optimizing experimental target choice, we have performed a comparative study of more than twenty materials for their discovery reach via single phonon excitations~\cite{Griffin:2019mvc,Coskuner:2021qxo} and compiled the results in an online repository~\cite{demo}. 
Fig.~\ref{fig:phonon} shows a representative subset of these results, for a benchmark DM model with a light dark photon mediator. 
We see that the well-motivated freeze-in benchmark is very well within reach of a table-top experiment with kg-yr exposure. 
See Ref.~\cite{Coskuner:2021qxo} for details.

Also important, but more challenging, is to characterize the multi-phonon response. 
This is the interpolating regime between single phonon excitations and nuclear recoils, and allows for discovery before single phonon sensitivity is achieved.
There have been initial investigations based on EFT techniques and oscillator models~\cite{Campbell-Deem:2019hdx,Kahn:2020fef,Knapen:2020aky}, and more sophisticated DFT calculations are underway. 

In addition to phonons, magnons -- quanta of collective spin excitations -- have also been proposed for detecting DM interactions with spin degrees of freedom~\cite{Trickle:2019ovy,Mitridate:2020kly}. 
This is closely related to the idea of the QUAX experiment~\cite{Ruoso:2015ytk,Barbieri:2016vwg,Crescini:2018qrz,Alesini:2019ajt,QUAX:2020adt}, which aims to detect axion DM induced classical spin waves.
Another recent proposal is to use antiferromagnetic topological insulators that host axion-like quasiparticles with tunable gap for axion DM detection~\cite{Marsh:2018dlj,Schutte-Engel:2021bqm,Chigusa:2021mci}.

There is a general EFT framework for analyzing collective excitations induced by DM scattering~\cite{Trickle:2020oki}. 
Starting from a particle model of DM, we can first match it onto an EFT containing non-relativistic operators coupling the DM to Standard Model particles $\psi=p,n,e$ (proton, neutron, electron). 
The list of operators form an extended basis compared to the nuclear recoil case~\cite{Fitzpatrick:2012ix,Cirelli:2013ufw,Anand:2013yka,Gresham:2014vja,Anand:2014kea,DelNobile:2018dfg}, due to in-medium violation of Galilean invariance. 
These operators then induce a set of crystal responses; at leading order in the long wavelength limit, these include couplings to particle numbers $\langle N_\psi\rangle$, spins $\langle \vect{S}_\psi\rangle$, orbital angular momenta $\langle \vect{L}_\psi\rangle$ and tensorial spin-orbit couplings $\langle \vect{L}_\psi\otimes\vect{S}_\psi\rangle$ of each ion in the crystal lattice. 
All four types of crystal responses can lead to phonon excitations while $\langle \vect{S}_e\rangle$ and $\langle \vect{L}_e\rangle$ can also lead to magnon excitations in appropriately chosen targets. 
The EFT framework allows us to assess the discovery prospects of experiments like SPICE and HeRALD across a broad range of DM models, well beyond those having a simple spin-independent interaction. 
The EFT calculation of single phonon rate has been implemented in an open-source program \textsf{PhonoDark}~\cite{phonodark}.

Finally, studies of collective excitations for DM detection have also uncovered interesting sum rules~\cite{Lasenby:2021wsc} and selection rules~\cite{Cox:2019cod,Mitridate:2020kly}. 
We hope further investigations in these direction will provide both important consistency checks for theoretical calculations and useful guidance for experimental design.

\vspace{10pt}
\paragraph*{Acknowledgments.}
We thank Marco Bernardi, Hsiao-Yi Chen, Ahmet Coskuner, Sin\'ead Griffin, Thomas Harrelson and Katherine Inzani for fruitful collaborations on subjects discussed in this white paper. 
A.M., T.T.\ and K.Z.\ were supported by the U.S.\ Department of Energy, Office of Science, Office of High Energy Physics, under Award No.~DE-SC0021431, and the Quantum Information Science Enabled Discovery (QuantISED) for High Energy Physics (KA2401032).
K.Z.\ was also supported by a Simons Investigator Award. 
Z.Z.\ was supported by the U.S.\ Department of Energy under the grant DE-SC0011702.

\bibliographystyle{apsrev4-1}
\bibliography{snowmass_WP_bib}

\end{document}